\documentclass[aps,floatfix,prd,twocolumn,a4paper,10pt,citeautoscript,
preprintnumbers,superscriptaddress,showpacs]{revtex4-1} 

\usepackage[english]{babel}	
\usepackage{amsmath,amssymb,amsfonts} 
\usepackage{bbm}			

\usepackage{graphicx}		
\usepackage{graphics}		
\DeclareGraphicsExtensions{.pdf}	

\usepackage[colorlinks=true, pdfstartview=FitV,linkcolor=blue, 
			citecolor=blue, urlcolor=blue]{hyperref}
\renewcommand{\theequation}{\arabic{equation}}

\makeatletter
\newcommand*{\Equation}{\@ifstar\sEquation\oEquation}
\newcommand{\sEquation}[1]{\begin{equation*}#1\end{equation*}}
\newcommand{\oEquation}[2]{  \begin{equation}\label{#1}#2\end{equation} }
\makeatother
\newcommand{\Align}[2]{\begin{align}\label{#1}#2\end{align}}

\newcommand{\bs}{\boldsymbol}

\newcommand{\Figref}[1]{Fig.~\ref{#1}}
\newcommand{\Eqref}[1]{\eqref{#1}}

\newcommand{\groupU}[1]{U(#1)}   
\newcommand{\groupSU}[1]{SU(#1)} 
\newcommand{\groupZ}[1]{\mathbb{Z}_{#1}} 

\newcommand{\Grad}{\bs \nabla}

\newcommand{\Curl}{\bs \nabla\times}

\newcommand{\ez}{\bs e_z}

\newcommand{\D}{{\bs D}}
\newcommand{\A}{{\bs A}}
\newcommand{\B}{{\bs B}}

\newcommand{\F}{\mathcal{F}}
\newcommand{\G}{\mathcal{G}}

\begin{document}
\title{Vortex chains due to nonpairwise interactions and 
field-induced phase transitions between states with different 
broken symmetry in superconductors with competing order parameters
}
\author{Julien~Garaud}
\email{garaud.phys@gmail.com}
\affiliation{Department of Theoretical Physics, 
KTH-Royal Institute of Technology, Stockholm, SE-10691 Sweden}
\author{Egor~Babaev}
\affiliation{Department of Theoretical Physics, 
KTH-Royal Institute of Technology, Stockholm, SE-10691 Sweden}
\date{\today}

\begin{abstract}

We study superconductors with two order components and phase 
separation driven by intercomponent density-density interaction, 
focusing on the phase where only one condensate has non-zero 
ground-state density and a competing order parameter exists 
only in vortex cores. 
We demonstrate there, that multi-body intervortex interactions 
can be strongly non-pairwise, leading to some unusual vortex 
patterns in an external field, such as vortex pairs and vortex 
chains.
We demonstrate that, in external magnetic field, such a 
system undergoes a field-driven phase transition from 
(broken) $U(1)$ to (broken) $U(1)\times U(1)$ symmetries, 
when the subdominant order parameter in the vortex cores 
acquires global coherence.
Observation of these characteristic ordering patterns in 
surface probes may signal the presence of a subdominant 
condensate in the vortex core. 

\end{abstract}

\maketitle


\section{Introduction}
 
The unusual magnetic response that originates in multi-scale 
inter-vortex interactions recently attracted substantial 
interest in the framework of multi-component superconductivity. 
The interest was sparked by the observations of vortex aggregates 
in the two-band superconductor MgB$_2$ \cite{Moshchalkov.Menghini.ea:09,
Nishio.Dao.ea:10,Dao.Chibotaru.ea:11,Li.Nishio.ea:11,
Gutierrez.Raes.ea:12}, multi-band iron pnictides 
Ba$($Fe$_{1-x}$Co$_x)_2$As$_2$ \cite{Luan.Auslaender.ea:10,
Kalisky.Kirtley.ea:11} and Ba$_{1-x}$K$_x$Fe$_2$As$_2$ 
\cite{Vinnikov.Artemova.ea:09}, as well as in spin triplet 
Sr$_2$RuO$_4$ \cite{Hicks.Kirtley.ea:10,Ray.Gibbs.ea:14}.
There, the existence of multiple coherence lengths may lead 
to multi-scale physics that can account for observation of 
vortex aggregates.
On the other hand, models of multi-component superconductivity
featuring bi-quadratic density-density interaction are currently 
discussed in the context of superconductors with pair density 
wave order \cite{Agterberg.Tsunetsugu:08,Berg.Fradkin.ea:09a}, 
and most recently in the context of interface superconductors 
such as SrTiO$_3$/LaAlO$_3$ \cite{Agterberg.Babaev.ea:14}.
Here we investigate the properties of topological defects in 
an immiscible phase of a two component model, where there is 
strong bi-quadratic interaction that penalizes coexistence of 
both superconducting condensates. We show that it features 
unusual multi-scale physics of the vortex matter where 
non-pairwise interactions are important.
This is modelled by a theory of two complex fields, that 
have a $\groupU{1}\times\groupU{1}$ symmetry. In the 
phase-separated regime, that occurs for strong bi-quadratic 
interaction, the ground-state spontaneously breaks only one 
of the $\groupU{1}$ of the symmetry of the theory.

In two-component superconductors, when both condensates have 
non-zero ground-state density, non-monotonic interactions 
can occur, due to competing inter-vortex interactions with 
different length scales \cite{Babaev.Speight:05,
Babaev.Carlstrom.ea:10,Carlstrom.Babaev.ea:11}. This typically 
leads to formation of vortex clusters surrounded by macroscopic 
regions of Meissner state \cite{Carlstrom.Garaud.ea:11}. 
Because it features properties of both type-1 and type-2 
superconductors, this regime is termed type-1.5. It
is a subject of ongoing studies, both experimental on MgB$_2$ 
\cite{Moshchalkov.Menghini.ea:09,Nishio.Dao.ea:10,Li.Nishio.ea:11,
Gutierrez.Raes.ea:12} and more recently in Sr$_2$RuO$_4$ 
\cite{Ray.Gibbs.ea:14} and theoretical studies of Ginzburg-Landau 
\cite{Babaev.Carlstrom.ea:10,Carlstrom.Babaev.ea:11,
Garaud.Agterberg.ea:12}, microscopic \cite{Silaev.Babaev:11} 
and effective point-particle models \cite{Drocco.Reichhardt.ea:13,
Varney.Sellin.ea:13}.

Here, we show that unusual multi-scale interaction arises 
in models of two-component superconductors with strong 
intercomponent bi-quadratic coupling that is repulsive. 
The bi-quadratic interaction penalizes coexistence of both 
condensates and above a given critical coupling they 
cannot coexist, so that one is completely suppressed. 
However, in the cores of vortices, this interaction 
is effectively much weaker and the suppressed component can 
locally condense. We demonstrate that the condensation in 
vortex cores leads to new unusual multi-scale, non-monotonic 
interactions between vortex matter, where non-pairwise forces 
are important (see also remark 
\footnote{
Because the Ginzburg-Landau theory is nonlinear, non-pairwise 
forces between vortices are generically present. However, 
in simple single-component systems, such forces do not affect 
structure formation. They merely decrease the two-body forces 
in dense vortex lattices, thus being responsible for crossover 
between known vortex solutions near $H_{c1}$ and lattice solutions 
$H_{c2}$ \cite{SaintJames,Chaves.Peeters.ea:11,Edstrom:13} 
~
}).  \nocite{SaintJames,Chaves.Peeters.ea:11,Edstrom:13}
Because it originates in multiple condensates with a particular 
hierarchy of the physical length scales, it is somewhat akin to 
the type-1.5 regime, but with the substantial difference here 
that only one condensate has non-zero ground-state density.

\begin{figure*}[!htb]
\hbox to \linewidth{ \hss
\includegraphics[width=0.75\linewidth]{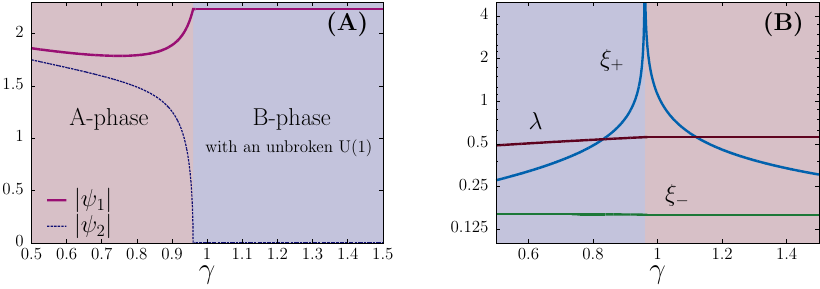}
\hss}
\vspace{-0.3cm}
\caption{
(Color online) -- 
ground-state properties of the model. The panels (A) 
and (B) respectively display ground-state densities 
and length scales (computed from the eigenvalues of 
the Hessian matrix \Eqref{Hessian}), when the intra 
condensate couplings are not equal $\alpha_1=-5$, 
$\alpha_2=-4.8$. $\beta_1=\beta_2=1$ and $e=0.8$. 
Depending on the strength of the bi-quadratic coupling 
$\gamma$, the ground-state corresponds to either the 
A-phase or the B-phase, as defined in Eqs.~\Eqref{Aphase} 
and \Eqref{Bphase}. One of the length scales, $\xi_+$, 
diverges at the critical value $\gamma_\star$ that 
separates both phases, while the other one, $\xi_-$ 
is always finite.
}
\label{Fig:PhaseDiag}
\end{figure*}

Below, we study the two-component Ginzburg-Landau model 
where intercomponent density-density interaction can be 
strong enough to completely suppress one of the condensates, 
in the ground-state. We characterize the different possible 
ground-state phases of that model and the associated length 
scales. Finally, we numerically investigate the properties 
of vortices within the phase above a critical density-density 
coupling, where both components cannot coexist. There we 
demonstrate the existence of the above mentioned regime 
where intervortex interactions are non-monotonic, and where 
multi-body forces are important. Unlike the type-1.5 regime 
where vortices typically aggregate into clusters \cite{
Babaev.Carlstrom.ea:10,Carlstrom.Babaev.ea:11,Garaud.Agterberg.ea:12}, 
vortices here tend to form chains and irregular structures.
Unlike chains forming in multi-scale systems with long-range 
repulsive interaction \cite{Malescio.Pellicane:03,Glaser:07,
Dinsmore.Dubin.ea:11,CostaCampos.Apolinario.ea:13,
McDermott.OlsonReichhardt.ea:14}, chains here originate 
in non-pairwise intervortex forces.


\section{The Model}

The Ginzburg-Landau model we consider here is a theory 
two complex fields $\psi_1$ and $\psi_2$ standing for two 
superconducting condensates. They interact together by 
their coupling to the vector potential of the magnetic field 
$\B=\Curl\A$, through the kinetic term $\D\equiv\Grad+ie\A$:
\Align{FreeEnergy}{
 \mathcal{F}= \frac{\B^2}{2}+&\sum_{a=1,2}\Big\{\frac{1}{2}|\D\psi_a|^2   	
+\alpha_a|\psi_a|^2+\frac{1}{2}\beta_a|\psi_a|^4 	\Big\} \nonumber\\
+&\gamma|\psi_1|^2|\psi_2|^2
\,.
}
Moreover, the condensates are directly coupled together by a 
bi-quadratic (density-density) interaction potential term when 
$\gamma\neq0$ and because the bi-quadratic interaction is 
repulsive, $\gamma>0$. For generic values of the parameters of 
the potential, $\alpha$, $\beta$ and $\gamma$'s, the theory 
has a $\groupU{1}\times\groupU{1}$ symmetry.
\footnote{
For special values of the parameters $\alpha_1=\alpha_2$ 
and $\beta_1=\beta_2$ the symmetry of the theory is 
enlarged to $\groupU{1}\times\groupU{1}\times\groupZ{2}$. 
If on top that $\gamma=\beta_1=\beta_2$, the theory has 
an even higher symmetry group: $\groupSU{2}$. These situations 
we do not consider here. We thus assume that 
$\alpha_1\neq\alpha_2$.
}

Depending on the relation between the parameters of 
the potential, two qualitatively different superconducting 
phases can be identified.
These are determined by the ground-state properties 
of the theory. Since the potential depends on the fields 
moduli only, the ground-state is the state with constant 
densities of the superconducting condensates $|\psi_a|=u_a$ 
and where the vector potential is a pure gauge 
($\A=\Grad\chi$ for arbitrary $\chi$) that can consistently 
chosen to be zero. The extrema of the potential, are given 
by $\partial V/\partial|\psi_a|=0$ and the ground-state 
densities $u_a$ satisfy: 
\Equation{Extrema}{
\left\lbrace \begin{array}{c}
2\left(\alpha_1+\beta_1u_1^2+\gamma u_2^2\right)u_1 =0\,\phantom{.}\\
2\left(\alpha_2+\beta_2u_2^2+\gamma u_1^2\right)u_2 =0\,.
\end{array}\right. 
}
For the extrema to be stable (minima), the eigenvalues 
of the Hessian matrix $\mathcal{H}=
\partial^2 V/\partial|\psi_a|\partial|\psi_b|$ must be 
positive. Here the Hessian matrix reads 
\Equation{Hessian}{
\mathcal{H}= 2\left(\begin{array}{c c}
\alpha_1+3\beta_1u_1^2+\gamma u_2^2	& 2\gamma u_1u_2 \\
2\gamma u_1u_2		&\alpha_2+3\beta_2u_2^2+\gamma u_1^2
\end{array}\right)\,. 
}
Apart from the normal state ($u_1=u_2=0$), there are 
two qualitatively different solutions of \Eqref{Extrema}: 
the A-phase (miscible) for which both condensates have 
non-zero ground-state density ($u_1,u_2\neq0$), and the 
B-phase (immiscible) for which only one condensate has 
non-zero ground-state density: either $u_1\neq0$ and 
$u_2=0$ or $u_1=0$ and $u_2\neq0$. Assuming that 
$\alpha_a<0$ and $\beta_a>0$, the qualitatively 
different stable phases determined by \Eqref{Extrema} 
and \Eqref{Hessian} are
\Align{Aphase}{
&\textbf{A-phase:~} (u_1^2,u_2^2)=
\left(\frac{\alpha_2\gamma-\alpha_1\beta_2}{\beta_1\beta_2-\gamma^2}
\,,\frac{\alpha_1\gamma-\alpha_2\beta_1}{\beta_1\beta_2-\gamma^2}	\right)	 
\\
&\text{if~}	\beta_1\beta_2>\gamma^2\,,~	\alpha_2\gamma-\alpha_1\beta_2>0 
\text{~and~}\alpha_1\gamma-\alpha_2\beta_1>0\,. \nonumber\\
&\textbf{B-phase:~} (u_1^2,u_2^2)=
\left(\frac{-\alpha_1}{\beta_1}\,,0\right)
~~\text{or}~~
\left(0,\frac{-\alpha_2}{\beta_2}\right)
\label{Bphase}	\\
&\text{if~}\alpha_2\beta_1-\alpha_1\gamma>0
~~\text{or}~~\alpha_1\beta_2-\alpha_2\gamma>0
\,. \nonumber
}
Clearly, to understand properties of the B-phase it 
is enough to consider only the first case where $u_1\neq0$ 
and $u_2=0$, as the case $u_2\neq0$ and $u_1=0$ can 
straightforwardly be obtained from the first one. Note that 
we disregard the possibility of having one positive 
$\alpha_a$. For both $\alpha_a>0$, the ground-state 
is the normal state $u_1=u_2=0$.
The ground-state in the A-phase spontaneously breaks 
the $\groupU{1}\times\groupU{1}$ symmetry. In the B-phase, 
only one of the $\groupU{1}$'s is spontaneously broken 
while the other, associated to the suppressed condensate, 
remains unbroken.

In this work, we are primarily interested in the properties of 
the B-phase \Eqref{Bphase}, in the vicinity of the phase transition 
between A- and B- phases. A convenient parametrization to understand 
this transition is to investigate the role of the bi-quadratic 
coupling $\gamma$. As shown in \Figref{Fig:PhaseDiag}, for fixed 
values of $\alpha_a$ and $\beta_a$, the bi-quadratic coupling 
$\gamma$ can be used to parametrize the transition between the 
two phases. The length scales $\xi_\pm$ are defined from the 
eigenvalues $m_\pm^2$ of the Hessian \Eqref{Hessian} as 
$\xi_\pm=1/m_\mp$, while the penetration depth is 
$\lambda=1/e\sqrt{u_1^2+u_2^2}$. Here $m^2_+$ stands for the largest 
eigenvalue of the Hessian and $m^2_-$ the smallest. The relation 
between the Hessian matrix and the length scales can be heuristically 
understood as follows. The Hessian matrix contains the informations 
about the stability of the ground-state and thus how it recovers 
from a small perturbation. It is important to understand that 
$\xi_\pm$ corresponds to hybridized modes and cannot be attributed 
to a given condensate separately. That is, $m^2_\pm$ are the decay 
rates of a linear combination of $\psi_1$ and $\psi_2$.
Long-range intervortex interaction is controlled by the masses of 
normal modes. The linearized theory yields the following long-range 
intervortex interaction \cite{Carlstrom.Babaev.ea:11}:
\Equation{asymp}{
V=q_\lambda K_0(r/\lambda) - q_-K_0(r/\xi_-) -  q_+K_0(r/\xi_+) \,,
}
where $K_0$ is the modified Bessel function of the second kind and the 
coefficients $q_\lambda$ and $q_\pm$ are determined by nonlinearities.
Here the first term describes the repulsion driven by current-current 
and magnetic interactions, while the second and third terms describe 
density-field-driven interactions. 

Single component superconductors are classified into type-1/type-2 
when the penetration depth $\lambda$ is smaller/larger than the 
coherence length $\xi$. From this, the vortex interactions are 
attractive in type-1 because long range interaction is mediated by 
core-core interactions. On the other hand, it is repulsive for type-2, 
due to current-current interactions that range with $\lambda$.
In two-component superconductors, such a classification is not directly 
applicable because of the existence of multiple length scales $\xi_\pm$. 
In particular, if the penetration depth is an intermediate length scale, 
$\xi_-<\lambda<\xi_+$, it, under certain conditions, leads to non-monotonic 
interactions that are long-range attractive and short-range repulsive 
\cite{Babaev.Speight:05,Carlstrom.Babaev.ea:11}. This can result in the 
formation of vortex clusters surrounded by macroscopic regions of 
Meissner state \cite{Carlstrom.Garaud.ea:11}. 
This phase is coined type-1.5 and observation of clusters were reported 
from measurements in clean MgB$_2$ \cite{Moshchalkov.Menghini.ea:09,
Moshchalkov.Menghini.ea:09,Nishio.Dao.ea:10} and in Sr$_2$RuO$_4$ 
\cite{Ray.Gibbs.ea:14} samples.

When increasing $\gamma$, toward the critical value 
$\gamma_\star=\alpha_2\beta_1/\alpha_1$ that separates A- and 
B- phases, the disparity in densities becomes more important. 
This is accompanied with the increase of the largest length 
scale, $\xi_+$. At $\gamma_\star$ this length scale diverges, 
while all the other length scales remain finite. 
In the A-phase, where both condensates have non-zero ground-state 
density, elementary topological excitations are vortices with 
winding in either condensate. These carry a fraction of the flux 
quantum, but finiteness of the energy imposes that they form a 
bound state that has phase winding in both condensates and that 
carries integer flux quantum. The most simple version of such a 
bound state is  to have vortices in both condensates and that 
they superimpose. 
However, solutions where vortices do not coincide can exist and 
be preferred energetically. It has recently been argued that such 
topological defects, characterized by an additional topological 
invariant, could be realized in interface superconductors, such 
as SrTiO$_3$/LaAlO$_3$ \cite{Agterberg.Babaev.ea:14}.
If $\lambda$ is not the smallest length scale (i.e. not a type-1 
regime), then there always exists a regime, in the vicinity of 
$\gamma_\star$, where the penetration depth is an intermediate 
length scale: $\xi_-<\lambda<\xi_+$. In the A-phase, this length 
scale hierarchy is known to be a necessary condition for the 
non-monotonic vortex interaction \cite{Babaev.Carlstrom.ea:10}. 
Clearly, this is realized close to $\gamma_\star$, see 
\Figref{Fig:PhaseDiag}.


\section{Evidences for strong non-pairwise inter vortex forces}

Here our main interest are the properties of the B-phase, 
in particular in the vicinity of $\gamma_\star$. In contrast 
to the above mentioned type-1.5 regime of the A-phase, the 
topological excitations in the B-phase are vortices that 
have core in $\psi_1$ only. Away from vortex cores, the 
fields recover their ground-state values and thus only 
$\psi_1$ can contribute to the flux quantization.

\begin{figure}[!htb]
\hbox to \linewidth{ \hss
\includegraphics[width=\linewidth]{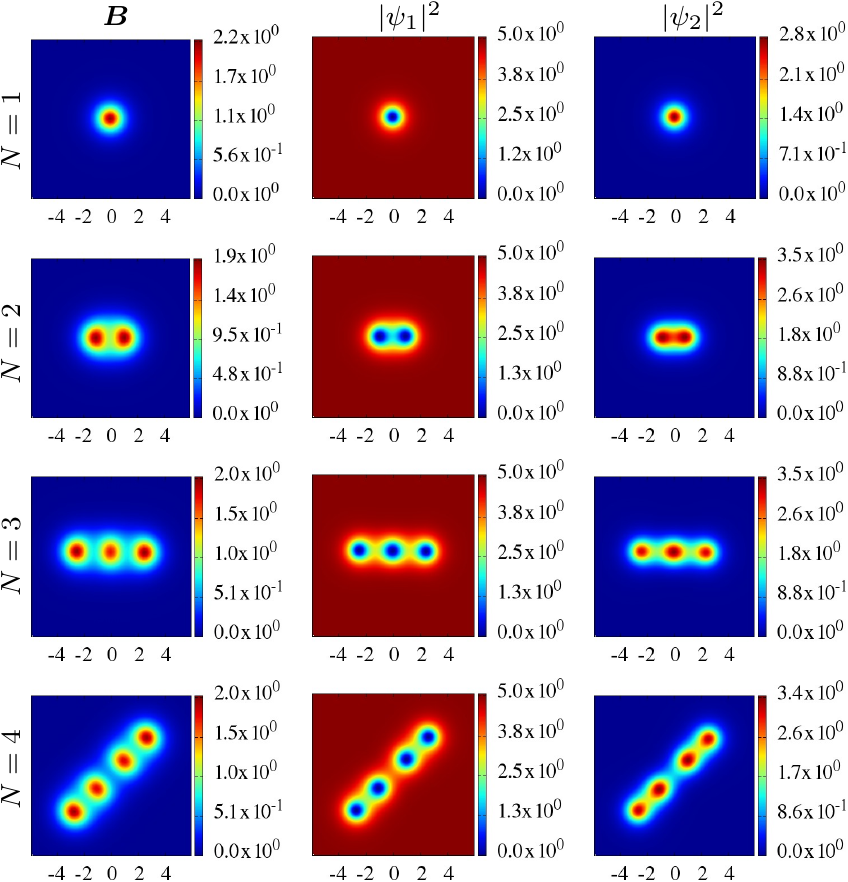}
\hss}
\vspace{-0.3cm}
\caption{
(Color online) -- 
Vortex solutions in the B-phase of \Figref{Fig:PhaseDiag}, for the 
coupling constant of the bi-quadratic interaction $\gamma=1.0$. 
The first column displays the magnetic field, while the second and 
third columns show $|\psi_1|^2$ and $|\psi_2|^2$, respectively. The 
lines show configurations carrying $N=1$, $2$, $3$, and $4$ flux 
quanta, respectively. 
In the B-phase, only $\psi_1$ has non-zero ground-state density, 
because the bi-quadratic coupling is too strong to allow coexistence 
of both condensates. Thus only $\psi_1$ forms vortices, while $\psi_2$ 
is zero everywhere except in vortex cores. 
As expected from the length scales considerations, intervortex 
interaction is non-monotonic and vortices stand at a preferred 
distance, see second line. 
For a larger number of flux quanta (third and fourth line), vortices 
form straight chains. This contrasts with the two-body picture that 
would predict formation of compact clusters. The chain-like structures 
thus signal existence of strong non-pairwise forces between vortices.
We should remark that the simulations are performed on a domain 
that is large enough, so that the vortices do not interact with 
boundaries. The plots show  only a small fraction of the numerical grid.
}
\label{Fig:PhaseSep}
\end{figure}

\begin{figure}[!htb]
\hbox to \linewidth{ \hss
\includegraphics[width=\linewidth]{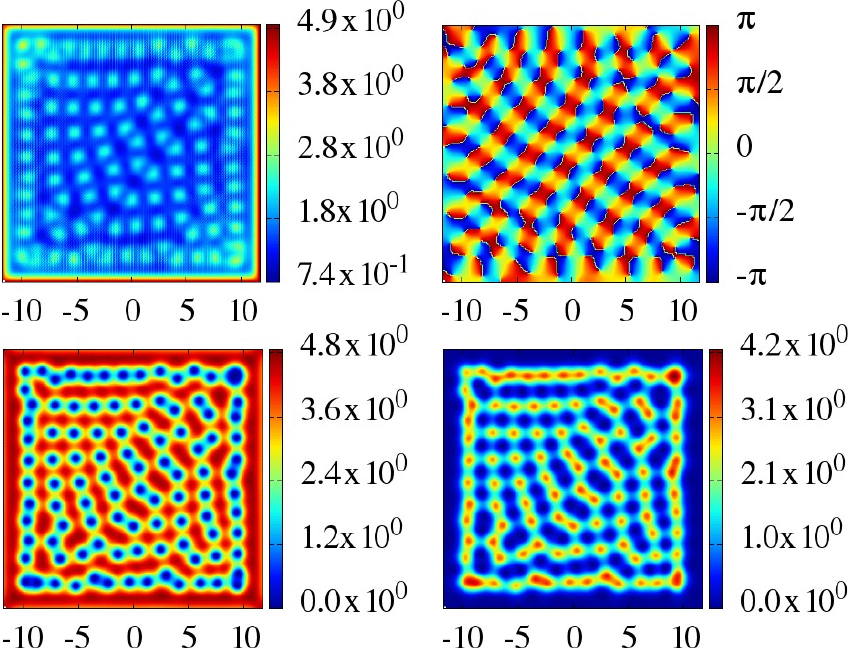}
\hss}
\vspace{-0.3cm}
\caption{
(Color online) -- 
The parameters are the same as in \Figref{Fig:PhaseSep}. The panels 
on the first row display the magnetic field and the phase difference 
$\varphi_{12}=\varphi_2-\varphi_1$. The second line shows the 
densities $|\psi_1|^2$ and $|\psi_2|^2$, respectively. 
Note that this configuration is not a true ground-state in 
external field, but is a very stable state. Here, the tendency 
to form chains competes with finite size effects, resulting in 
very irregular pattern for vortices. Due to the presence of 
multi-body forces, obtaining true ground-state in the simulations 
of magnetization processes for systems of these sizes turns out 
to be very difficult (For a discussion of glassiness arising
from the non-pairwise forces see \cite{Sellin.Babaev:13}). 
This suggests that the shown patterns should also be physically 
representative for experimental situations in such systems. 
}
\label{Fig:Applied1}
\end{figure}

To investigate the properties of topological excitations and their 
interactions, we numerically minimize the free energy \Eqref{FreeEnergy} 
within a finite element framework \cite{Hecht:12}. That is, for a 
given choice of parameters, a starting configuration with desired 
winding is created and the energy is then minimized with a non-linear 
conjugate gradient algorithm. For detailed discussion on the numerical 
methods, see for example appendix in Ref.~\onlinecite{Garaud.Babaev:14a}. 
In the B-phase only the condensate $\psi_1$, has non-zero 
ground-state density and thus only $\psi_1$, has vortex excitations. 
Since the component $\psi_1$ vanishes at the vortex core, it can be 
beneficial for the suppressed component $\psi_2$ to assume non-zero 
density in the cores of vortices. A similar mechanism of condensation 
in vortex cores was also discussed in the context of cosmic strings
\cite{Witten:85a}. Minimizing the free energy 
\Eqref{FreeEnergy} for an initial configuration carrying a 
single flux quantum relaxes to such a vortex state, see first line 
in \Figref{Fig:PhaseSep}.
The condensate $\psi_2$ that lives inside the vortex cores is 
gradually suppressed where the other condensate $\psi_1$ recovers 
toward its ground-state density. The rate at which $\psi_2$ 
recovers is determined by the fundamental length scales $\xi_\pm$ 
of the theory. Because the modes are hybridized, the length scales 
associated with the recovery of $\psi_1$ and the decay of $\psi_2$ 
are not independent.

In the B-phase, in the vicinity of $\gamma_\star$, the length 
scales satisfy the necessary condition for non-monotonic 
interactions. Indeed, as shown on the second line of 
\Figref{Fig:PhaseSep}, interactions between two vortices can 
also be non-monotonic in the B-phase, even if only one condensate 
has non-zero ground-state density. There, in agreement with 
the linear theory \Eqref{asymp}, pairwise interaction between 
vortices is long range attractive due to the largest hybridized 
density mode and short range repulsive due to current-current 
interactions. It results in a preferred distance at which vortices 
minimize their interaction energy by forming a vortex pair.
Based on these observations, natural expectation from the two-body 
interactions is that states with more than two vortices will form 
compact clusters inside which vortices tend to have triangular 
arrangement \cite{Carlstrom.Garaud.ea:11}.  
However because it is a non-linear problem, interactions between 
vortices can become more complicated, beyond the linear approximation. 
In particular, from studies  of point particle effective models 
\cite{Sellin.Babaev:13}, it follows that strong non-pairwise 
interactions can dramatically affect structure formation, resulting 
in stripe, gossamer, and glass phases.

The configurations for few isolated vortices displayed in 
\Figref{Fig:PhaseSep}, show chain organization of vortices. 
This indicates that there are non-monotonic interactions, but 
also that there are strong multi-body forces. Indeed the two-body 
picture would naively lead to conclude that many vortices would 
organize in a compact cluster. Because the theory \Eqref{FreeEnergy} 
is completely isotropic, the line-like organization can originate 
only in complicated interactions.
This poses the question of the response of the system to an external 
field. 
At elevated external field, vortex matter usually forms lattices 
(hexagonal, square, etc). Since the low field results indicate 
strong non-pairwise forces, the question arises if these have 
a substantial influence at elevated fields. To sort this out, we 
investigate the response in an external field ${\bs H}=H_z\ez$, 
perpendicular to the plane. For this, the Gibbs free energy 
$\G=\F-\B\cdot{\bs H}$ is minimized, with requiring that 
$\Curl\A={\bs H}$ on the boundary (see e.g discussion in appendix 
of Ref.~\onlinecite{Garaud.Babaev:14a}).
As shown in \Figref{Fig:Applied1}, the typical response in 
external field  shows a long-living irregular vortex structure.
For example, similar simulations, but in the A-phase, show very 
regular square lattices \cite{Garaud.Sellin.ea:14}. We show such 
a lattice in the Appendix~\ref{Appendix}.

There is a tendency here to form chains, but this tendency 
competes with the increased importance of current-current 
interactions in the relatively dense vortex matter. 
Note that the non-pairwise forces, when strong enough, 
typically promote metastable or long-living disordered states.
Also, when minimizing the Gibbs free energy with the condition
that $\Curl\A={\bs H}$ on the boundary, the interaction energy
between vortices is minimized not independently from the 
interaction with the Meissner currents on the boundary. 
Such finite size effects, play as well a role in having 
imperfect lattices.

Observe that it was demonstrated earlier, that in type-1.5 systems, 
multibody forces can aid formation of vortex chains for dynamic 
and entropic reasons \cite{Carlstrom.Garaud.ea:11}. However, here 
the non-pairwise forces are clearly much stronger, as chains form 
as ground-state solutions in low fields, see \Figref{Fig:PhaseSep}.
Note also that the chains and vortex dimers forming here 
originate in non-pairwise interactions and not because of 
pairwise interactions with multiple repulsive length scales 
\cite{Malescio.Pellicane:03,Glaser:07,OlsonReichhardt.Reichhardt.ea:10}.
They should also not be confused with vortex chains predicted for 
multilayer structures, where they originate in stray field 
that lead to long-range repulsive interaction \cite{Varney.Sellin.ea:13,
Komendova.Milosevic.ea:13}.

\section{Inducing state with different broken-symmetry by applied field}

For isolated vortices in $\psi_1$, the other component 
$\psi_2$ develops non-zero amplitude in the vortex core. 
However, as shown in \Figref{Fig:PhaseSep}, $\psi_2$
is asymptotically suppressed and thus it has has no 
phase winding. 
As mentioned in the introduction, in this state the system 
breaks only one $U(1)$ symmetry. In high external 
field, there is a large density of vortices and on average 
$|\psi_2|$ becomes non zero. There, the areas with non-zero 
$|\psi_2|$ get interconnected across the whole system 
and thus the system thus undergoes a phase transition to 
a state that breaks the $U(1)\times U(1)$ symmetry. 
By saying that the system breaks $U(1)\times U(1)$ symmetry 
in an external field we assume a robust vortex structure, 
we do not consider here vortex liquids. The interconnection 
of $\psi_2$ across the whole sample is signalled by a change 
in the phase winding pattern. If two condensates have non-zero 
density, phase winding in only one condensate gives a 
logarithmically divergent contribution to the energy 
\cite{Babaev:02}. As a result, it is energetically beneficial 
for the component $\psi_2$ to form vortices as well. This is 
in strong contrast with the results for isolated vortices. 
The breakdown of the $U(1)$ symmetry associated with the 
condensate $\psi_2$, and the corresponding formation of vortices 
can be seen from phase difference $\varphi_{12}=\varphi_2-\varphi_1$ 
shown in the upper right panel in \Figref{Fig:Applied1}. 
There, the dipole-like structure of $\varphi_{12}$ shows 
the existence of phase winding in both condensates but 
around different points. This unambiguously signals that 
both condensates have the same total phase winding and thus 
$\groupU{1}\times\groupU{1}$ symmetry-broken state.


\section{Metastable multi-quanta solutions}

When $\gamma$ becomes large enough as compared to $\gamma_\star$, 
condensation of $\psi_2$ in the vortex core becomes less important.
As a result, deeper in the B-phase, individual vortices show no 
condensation of $\psi_2$ in the core (see also remark 
\footnote{
Recently the problem of the conditions of appearance of the 
subdominant component condensation in  a single vortex core 
attracted interest in the context of cold atoms 
\cite{Catelani.Yuzbashyan:10}. However the results from such 
electrically neutral systems are not straightforwardly 
applicable to the charged systems because of the difference 
between power-law (in the context of neutral systems) vs 
exponential (in charged systems) vortex core localization. 
Note also the difference in the expressions for the coherence 
lengths between this and the above quoted paper.
}). \nocite{Catelani.Yuzbashyan:10}
\begin{figure}[!htb]
\hbox to \linewidth{ \hss
\includegraphics[width=\linewidth]{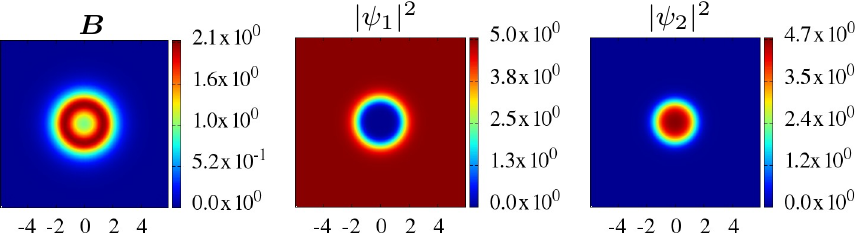}
\hss}
\vspace{-0.3cm}
\caption{
(Color online) -- 
Meta-stable solution, deep into the B-phase. This is a localized 
configuration that carries four flux quanta, for the same parameters 
as in \Figref{Fig:PhaseSep} except that $\gamma=1.2$. 
This object carries multiple flux quanta, despite not being in
a type-1 regime. It is made of a large central region of the 
condensate $\psi_2$ where $\psi_1=0$, embedded in a domain where 
$\psi_2=0$. The magnetic flux is screened by $\psi_1$ outside the 
vortex, while $\psi_2$ is responsible for screening inside. As a 
result, the magnetic flux is localized on a cylindrical shell around 
the vortex and resembles a pipe. 
}
\label{Fig:MetaStable}
\end{figure}
Moreover, deep into the B-phase, 
$\lambda$ becomes the largest length scale, and the interaction 
between vortices becomes long-range repulsive. Since this follows 
from asymptotic analysis, this holds sufficiently far from 
the vortex core. However, it does not preclude more involved 
interactions at shorter ranges. We computed vortex solutions as 
in \Figref{Fig:PhaseSep}, but deeper in the B-phase 
($\gamma=1.2, 1.4,\cdots$). There, we find that indeed, isolated 
vortices are preferred over vortex bound states. Nevertheless, 
we could find a special kind of metastable bound states of 
vortices. Namely, we found configuration carrying $N$ flux quanta 
whose energy $E(N)$ is larger than the one of $N$ isolated vortices: 
$E(N)>NE(N=1)$. These configurations are thus local minima of 
the energy functional and, for the parameters which we considered, 
they differ by less than 5 percent from isolated vortices. 
Such a meta-stable state is shown in \Figref{Fig:MetaStable}.
Being obtained through energy minimization, it is stable to small 
perturbations and depends on the starting configuration. Namely, 
if the starting configuration is in the attractive basin of the 
local minimum, it will converge to the local minimum. Typically 
if the starting configuration consists of dense packing of vortices, 
then it may lead to the meta-stable bound state. 
\begin{figure}[!htb]
\hbox to \linewidth{ \hss
\includegraphics[width=\linewidth]{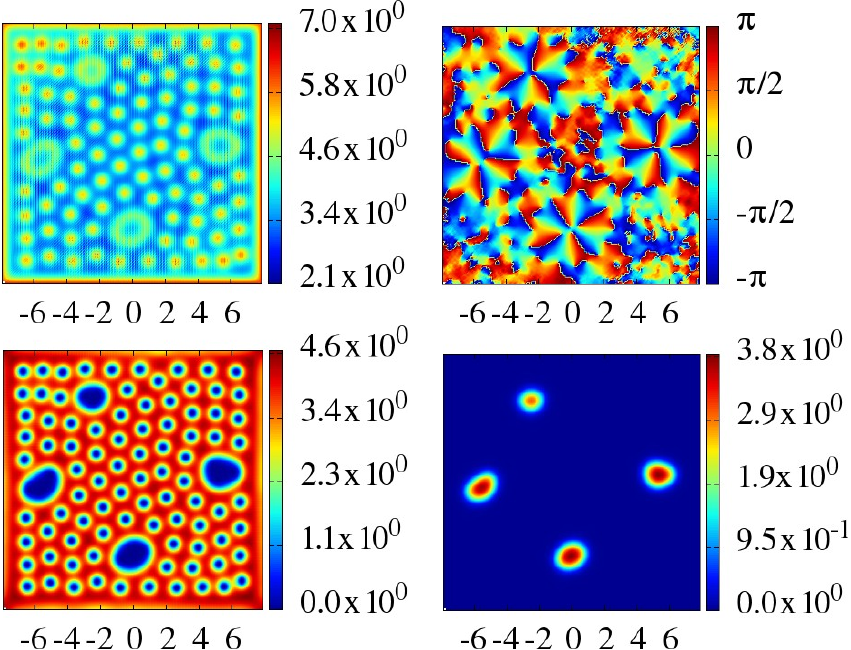}
\hss}
\vspace{-0.3cm}
\caption{
(Color online) -- 
Solution in an external field for the same parameters as 
in \Figref{Fig:Applied1}, but stronger bi-quadratic coupling 
$\gamma=1.5$. These parameters for the potential set the 
system deep into the B-phase where the penetration depth 
is the largest length scale. Thus it should behave as an 
ordinary type-2 system.
In such a regime, preferred solutions are isolated Abrikosov 
vortices. However, there also exist meta-stable states as 
the one shown in \Figref{Fig:MetaStable}.
The meta-stable bound state of vortices appears as an 
inclusion of a domain where $\psi_2$ condenses. Because 
these are surrounded by vortices exerting some pressure, 
in practice they do not decay into ordinary vortices.
}
\label{Fig:Applied2}
\end{figure}
The meta-stable state shown in \Figref{Fig:MetaStable} are lumps 
where $\psi_2$ is non zero, despite that, from an energetic 
viewpoint it should be suppressed.
There the magnetic flux is screened by $\psi_1$ 
outside the vortex, while $\psi_2$ is responsible for screening 
inside. As a result, the magnetic flux is localized on a cylindrical 
shell around the vortex and resembles a pipe. In different 
systems, similar pipe-like configurations can actually appear as 
true stable states for the special case where $\alpha_1=\alpha_2$ 
and $\beta_1=\beta_2$. This was recently investigated in a separate 
work \cite{Garaud.Babaev:14b}. Also pipe-like vortices were discussed 
in the Bogomol'nyi regime of $SU(2)$ theory where additionally 
$\gamma=\beta_1=\beta_2$ \cite{Chernodub.Nedelin:10}. 
There, the pipe-like solutions feature both properties of 
vortices and domain walls.
The remarkable feature of the pipe-like vortices in this regime, 
is that here the model does not have topological domain walls 
solutions. This makes it distinct from the other models that support 
meta-stable bound states of vortices due to existence of a broken 
$\groupZ{2}$ symmetry \cite{Garaud.Babaev:14b,Garaud.Carlstrom.ea:13,
Garaud.Carlstrom.ea:11}.

According to the asymptotics, intervortex interactions are long 
range repulsive. The attractive channel is activated only at 
shorter range. This means that when there are many vortices, 
relatively close to each other, they may form the bound states 
similar to the one displayed in \Figref{Fig:MetaStable}, because 
of the ``pressure'' of other vortices. Such a situation is likely 
to occur in external field and it may result in coexistence of 
single vortices and bound vortices. As shown in \Figref{Fig:Applied2}, 
this indeed happens, despite that the parameters are deep into the 
B-phase. Note that the energy difference and the stability of 
bound vortices depends on all parameters of the free energy. More 
precisely, when the difference between $\alpha_a$ is important 
then the meta-stable solution does not form anymore in our simulations. 
Thus, the coexistence of bound vortices and usual vortices is not 
a universal feature and needs both condensates to have parameters 
with rather similar values.

\begin{figure}[!htb]
\hbox to \linewidth{ \hss
\includegraphics[width=\linewidth]{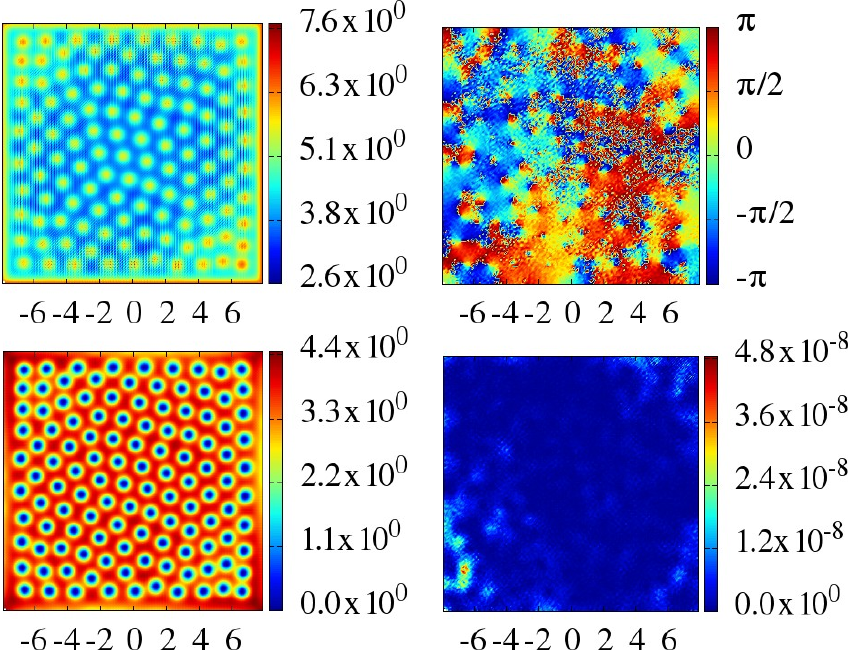}
\hss}
\vspace{-0.3cm}
\caption{
(Color online) -- 
Solution in an external field for the same parameters 
as in \Figref{Fig:Applied1}, but stronger bi-quadratic 
coupling $\gamma=1.6$. These parameters set the system 
deep into the B-phase where the penetration depth is no 
longer an intermediate length scale. Thus, it behaves 
as an ordinary type-2 system.
There, vortices have no condensation of $\psi_2$ inside 
the core, as can be seen from the last panel. Vortices 
in $\psi_1$ behave as regular Abrikosov vortices and try 
to arrange as a triangular lattice. Finite size effects 
and interaction with Meissner current deform the lattice, 
so that it is not really triangular. Note that since $\psi_2$ 
is zero (up to numerical precision), the phase difference 
$\varphi_{12}$ is reduced to numerical noise.
}
\label{Fig:Applied3}
\end{figure}

In our simulation of the model, the creation of the 
pipe-like meta-stable states was very history dependent. 
However, if they are created at all, it may be very difficult 
to destroy them.  That is, if isolated, pipe-like vortices 
are only meta-stable and may be very sensitive to small 
perturbations that can trigger decay into ordinary vortices. 
However, when surrounded by vortices, the decay channel may 
be different. Indeed, because it is type-2, vortices interact 
repulsively and they exert some pressure on the lump whose 
decay may thus be  more difficult. We show in 
\Figref{Fig:Applied2}, that this is indeed the case that 
in external field, deep into the B-phase, lumps coexist 
with vortices. Note that because their creation depends on 
past configurations, slowly ramping up the external field 
may make these more rare events. Deeper in the B-phase, 
pipe-like bound states are unstable and as shown in 
\Figref{Fig:Applied3}, there, only usual vortices $\psi_1$ 
exist and $\psi_2$ never condenses (up to numerical accuracy).


\section{Summary}

In this paper, we have investigated the physical properties of 
two-component Ginzburg-Landau models, with inequivalent components,
where bi-quadratic interactions penalize coexistence of both 
condensates. Above a critical coupling $\gamma_\star$, the condensates 
cannot coexist and only one preferred component can have non-zero 
ground-state density, thus breaking only one of the $\groupU{1}$ 
symmetries.  We have demonstrated that in a sufficiently strong 
magnetic field the second component nevertheless appears resulting 
in a phase transition where the (second) $U(1)$ symmetry is 
also broken. This kind of phase transition is by no means restricted 
to systems with $U(1)$ symmetry. It should also exist in other 
systems where different order parameters are localized at the core 
of topological defects.
Also we shown that under certain conditions such systems may 
form meta-stable states carrying multiple flux quanta distributed 
in a cylinder around the vortex, that resembles a pipe.  

Near the critical coupling $\gamma_\star$ one of the coherence 
lengths becomes the largest length scale. On the 
$\groupU{1}\times\groupU{1}$ side this results in the situation 
where the system cannot be a type-2 superconductor but be either 
of type-1 or type-1.5. In the later case one coherence length 
is larger and another is smaller than the magnetic field's 
penetration depth and the system vortices form clusters.  

Our main results pertain to the $U(1)$ ground-state, 
where both condensates are phase separated. There the simple 
picture from the two-body interactions fails to account for the 
structure of vortex bound states. Indeed, instead of forming 
vortex clusters as suggests the two-body picture, vortex chains 
are formed. Because the theory is fully isotropic, this 
can be interpreted as the hallmark of strong non-pairwise forces. 
These also affect the response in external field, where there 
is a clear tendency to form vortex chains. In a finite sample 
it results in rather irregular (metastable) vortex patterns with 
vortex dimers and vortex chains, as shown in \Figref{Fig:Applied1}. 
The result should hold for a variety of multicomponent models with 
competing order parameters. Thus observation of such vortex patterns 
may serve as an experimental hint for the presence of competing 
phases condensing in vortex cores. Interestingly the rather disordered 
vortex patterns are quite similar to those observed experimentally 
in iron-based superconductors \cite{Luan.Auslaender.ea:10,
Hicks.Kirtley.ea:10,Kalisky.Kirtley.ea:11}.
The richness of static and dynamic phases which can form in systems
with strong multi-body forces \cite{Sellin.Babaev:13,Sengupta.Sengupta.ea:10} 
calls for further investigation of vortex states in these models. 
In samples with disorder the pattern formation will be affected
by pinning which also calls for the investigation of its role. 
However, one can still expect prevalence of vortex pairs, in the 
presence of disorder.

\begin{acknowledgments}

We acknowledge   discussions with D.~F.~Agterberg and 
J.~Carlstr\"om.
The work was supported by the Knut and Alice Wallenberg Foundation 
through a Royal Swedish Academy of Sciences Fellowship, by the 
Swedish Research Council grants 642-2013-7837,  325-2009-7664. 
Part of the work was performed at University of Massachusetts, 
Amherst and supported  by the National Science Foundation under 
the CAREER Award DMR-0955902.
The computations were performed on resources provided by the 
Swedish National Infrastructure for Computing (SNIC) at National 
Supercomputer Center at Link\"oping, Sweden.
\end{acknowledgments}

\appendix
\setcounter{section}{0}
\setcounter{paragraph}{0}
\setcounter{equation}{0}
\renewcommand{\theequation}{\Alph{section}.\arabic{equation}}
\section{Vortex matter in the A-phase}
\label{Appendix}

\begin{figure}[!htb]
\vspace{0.2cm}
\hbox to \linewidth{ \hss
\includegraphics[width=.95\linewidth]{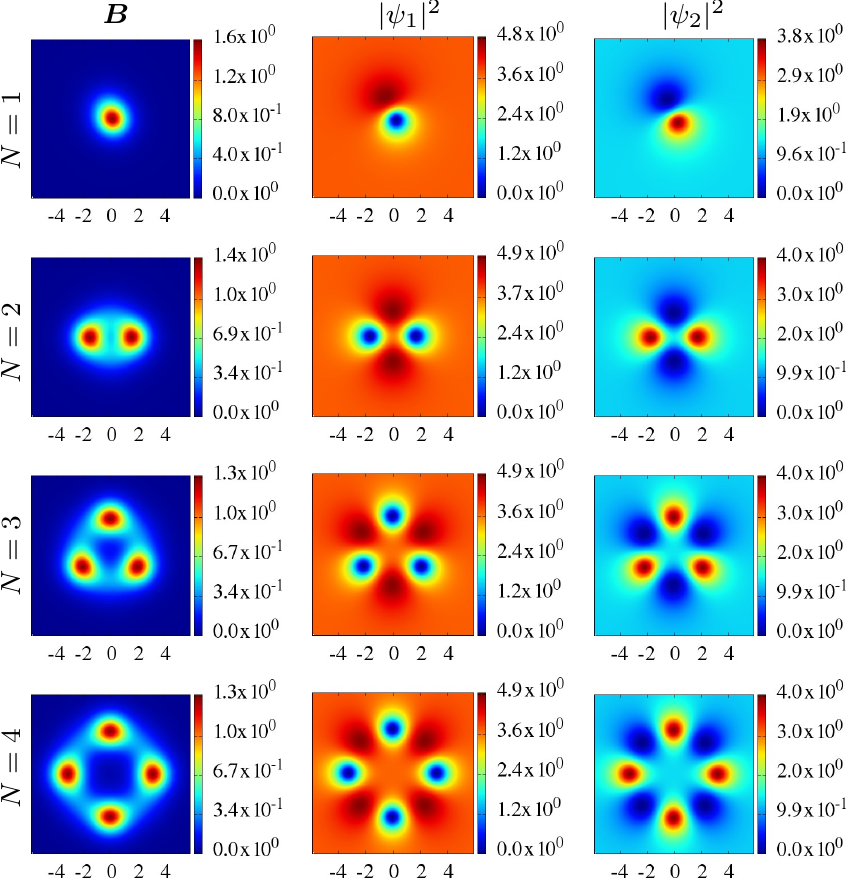}
\hss}
\vspace{-0.2cm}
\caption{
(Color online) -- 
Vortex solutions in the A-phase of the phase diagram 
\Figref{Fig:PhaseDiag}. There, the coupling constant of 
the bi-quadratic interaction is $\gamma=0.92$. Displayed 
quantities are the same as in \Figref{Fig:PhaseSep}. 
In the A-phase, where both components have non-zero 
ground-state densities, the bi-quadratic coupling makes 
it beneficial to split cores. This induces long range 
interaction between flux carrying defects through 
dipole interactions. This interaction is responsible 
for the binding of vortices.
}
\label{Fig:Dipoles}
\end{figure}
In the main body of the paper, we focus on vortex matter in B-phase 
where the bi-quadratic interactions are strong enough to segregate 
condensates. For completeness, in this appendix we provide additional 
materials that show the behavior of vortex matter in the A-phase for 
the model with these parameters (although it is not directly related 
to the main topic of the paper).

\begin{figure}[!htb]
\hbox to \linewidth{ \hss
\includegraphics[width=\linewidth]{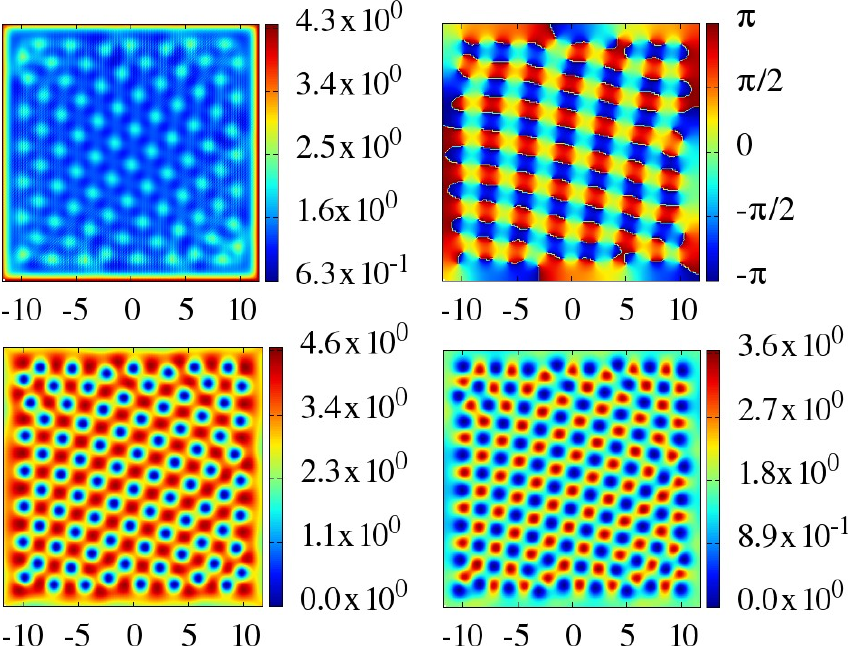}
\hss}
\vspace{-0.3cm}
\caption{
(Color online) -- 
Solution in an external field, for the applied field corresponding 
to $301$ flux quanta going through the sample's area. The parameters 
are the same as in \Figref{Fig:Dipoles} and displayed quantities are 
the same as in \Figref{Fig:Applied1}. Vortices in each condensate 
form square lattices that are translated from each other because of 
the bi-quadratic interaction. This results in a chequerboard pattern. 
Because of the disparity on ground-state densities, vortices in 
$\psi_2$ carry less flux than vortices in $\psi_1$. As a result the 
``brighter spots" of the magnetic field correspond to the vortices 
in $\psi_1$.
Note that the lattices are not perfect because of finite-size effects 
due to the interaction with Meissner currents and vortex entries at the 
boundaries.
}
\label{Fig:AppApplied}
\end{figure}

In the A-phase, both condensates have non-zero ground-state 
density. Thus, in order to have finite energy solutions both 
components must wind the same number of time. However, the cores 
do not necessarily have to overlap. Because of the bi-quadratic 
interaction, if the penetration depth is large enough, it is 
beneficial to split cores. As shown in \Figref{Fig:Dipoles}, 
the cores in $\psi_2$ do not superimpose with those in $\psi_1$. 
Core splitting in single vortices induces a dipolar interaction 
through the phase difference mode, that is long range. As can 
be seen in \Figref{Fig:Dipoles}, the long range dipolar forces 
heavily affect multiple vortex structure. This was discussed in 
slightly different models in Refs.~\onlinecite{Garaud.Sellin.ea:14,
Agterberg.Babaev.ea:14}.
The long range dipolar forces also heavily affect the magnetization 
process and the lattice solutions that are formed in high fields.
Indeed, in external field, vortices form a chequerboard pattern of 
two interlaced square lattices, as shown in \Figref{Fig:AppApplied}.

%

\end{document}